\documentclass[preprint2]{aastex}

\newcommand{\tar}{HD\,209458}
\newcommand{\tp}{HD\,209458\,b}
\newcommand{\nc}{HD\,210483}

\slugcomment{Accepted for publication in the Astrophysical Journal}
\shortauthors{Richardson et al.}

\received{2003 March 26}
\begin{document}

\title{Infrared Observations During the Secondary Eclipse of \tp \\
II. Strong Limits on the Infrared Spectrum Near 2.2 \micron}

\author{L. Jeremy Richardson\altaffilmark{1,2,3,4}, Drake
Deming\altaffilmark{1,2}, and Sara Seager\altaffilmark{5} }

\email{lee.richardson@colorado.edu}
\email{ddeming@pop600.gsfc.nasa.gov}
\email{seager@dtm.ciw.edu}

\altaffiltext{1}{Visiting Astronomer at the Infrared Telescope
Facility, which is
operated by the University of Hawaii under a cooperative agreement
with the National Aeronautics and Space Administration.}

\altaffiltext{2}{Planetary Systems Branch, Code 693, Goddard Space Flight Center, Greenbelt, MD 20771}
\altaffiltext{3}{Laboratory for Atmospheric and Space Physics,
University of Colorado, 1234 Innovation Drive, Boulder, CO 80303}
\altaffiltext{4}{Department of Physics, University of Colorado,
Boulder, CO 80309}
\altaffiltext{5}{Department of
Terrestrial Magnetism, Carnegie Institution of Washington, 5241 Broad
Branch Rd., NW, Washington, DC  20015}

\begin{abstract}
We report observations of the transiting extrasolar planet, \tp,
designed to detect the secondary eclipse.  We employ the method of
`occultation spectroscopy', which searches in combined light (star and planet)
for the disappearance and
reappearance of weak infrared spectral features due to the planet as it passes
behind the star and reappears.  Our observations cover two predicted
secondary eclipse events, and we
obtained 1036 individual spectra of the \tar\ system using the SpeX
instrument at the NASA IRTF in September 2001.  Our spectra 
extend from 1.9 to 4.2 \micron\ with a
resolution ($\lambda/\Delta \lambda$) of 1500.
We have searched for a continuum peak
near 2.2 \micron\ (caused by CO and H$_2$O absorption bands),
as predicted by some models of the planetary
atmosphere to be $\sim 6 \times 10^{-4}$ of the stellar flux, 
but no such peak is detected at a level of $\sim 3 \times
10^{-4}$ of the stellar flux.
Our results represent the strongest limits on the infrared spectrum
of the planet to date and carry significant implications for
understanding the planetary atmosphere. 
In particular, some models that assume the stellar irradiation is
re-radiated entirely on the sub-stellar hemisphere predict a flux peak
inconsistent with our observations.
Several physical mechanisms can
improve agreement with our observations, including the re-distribution of
heat by global circulation, a nearly isothermal atmosphere, and/or the
presence of a high cloud.\\
\end{abstract}


\section{INTRODUCTION}
The discovery of the first transiting extrasolar planet, \tp,
\citep{cb00,hmbv00} has led to several new observations designed to
characterize the physical properties of the planet.  These
observations have provided a determination of the planetary and stellar
radii and the true planetary mass \citep{bc01, cody}.  Furthermore, given
the geometry of the orbit, we are now beginning to learn
about the structure of the planet's atmosphere.  The atmosphere was
first probed conclusively by \citet{cb02}, who reported a detection of
the sodium doublet in transmission as the planet crossed in front of
the star.  Although typical models account for the strong stellar
irradiation and estimate that the effective temperature of
the planet is $T \simeq$ 1100--1800~K \citep[e.g.,][]{ss98, cb00},  no
measurements of the temperature
are available from actual observations of the planet.
An attempt to detect the reflected starlight from the planet during 
\emph{secondary eclipse} (i.e., the
time when the planet disappears behind the star) has recently been
reported \citep{kenworthy}, but
these observations were performed in the visible region and were limited by
instrumental and atmospheric effects.  
Pioneering infrared observations \citep{lucas02} did not achieve
sufficient sensitivity to detect realistic planetary models of \tp\
and similar extrasolar planet systems.
The infrared observations reported here, however, have sufficient
sensitivity to detect the thermal emission spectrum from \tp\ 
at the level predicted by several models for the planetary atmosphere.

In paper~I \citep{r1}, we applied the method of
occultation spectroscopy observations of the secondary eclipse from
the Very Large Telescope (VLT), and using this technique over a narrow
bandpass near 3.6~\micron,  we were able to
place limits on the abundance of methane in
the planetary atmosphere.  The most stringent limit applied only to an
exceptionally clear model atmosphere, and only for
certain values of the eclipse timing, which is uncertain by as much as
$\sim 30$~minutes, given the 1$\sigma$ error in the eccentricity of the orbit.
In this paper, we report a further attempt to detect the secondary
eclipse of \tp\ using the SpeX instrument at the NASA Infrared
Telescope Facility (IRTF).  
We took a different approach for these observations;
we used a broader wavelength range, in order to
look for the infrared continuum of the planet near 2.2~\micron.  By
observing the combined light from the star and planet, we
search for a change in the shape of the spectrum
as the planet disappears behind the star and later reemerges.
The nature of this technique makes our observations quite
sensitive to the temperature gradient in the planetary
atmosphere.

\section{OBSERVATIONS}
 
We obtained a total of four nights of data from the SpeX instrument
\citep{spex} at the
NASA IRTF located on Mauna Kea in Hawaii. 
These are UT 19,
20, 26, and 27 September 2001, and secondary eclipse events were
predicted for UT 20 and 27 September.

The SpeX instrument is a cross-dispersed echelle spectrometer capable
of imaging and spectroscopy in the wavelength region between 0.8 and
5.5 \micron\ with low to moderate spectral resolution
($\lambda / \Delta\lambda \lesssim 2500$).  We operated the instrument in 1.9 to 4.2
\micron\ mode using the 0.5 arcsec slit, giving a spectral resolution
of 1500; this mode gives nearly-continuous wavelength coverage in this
region, over six orders of the echelle.
The weather was excellent (low water vapor), and the seeing conditions were
good---less than 1.0 arcsec on all four nights.  Both eclipses
occurred within an hour of the transit of the star across the local
meridian, meaning that the eclipse was observed with the object
directly overhead.

The use of a comparison star is an important aspect of our observational
technique, and we selected \nc, a close spectral match to \tar\ and
nearby in the sky (see Table~1 in Paper~I).
The comparison
star was observed immediately following each observation of the \tar\
system.
We nodded the telescope between the `a' and
`b' positions on the slit to remove the terrestrial atmospheric 
background, and we recorded
spectra in the sequence `abba.'  A given spectrum at either slit
position was recorded with an integration time of 6~seconds per coadd
and a total of 3~coadds;
since the two objects were nearly the same visual magnitude, we used
the same integration time for observing both objects.
Typically we would record two successive `abba' sets of
\tar\ and then switch to the comparison star for two more `abba' sets,
giving approximately a 50\% duty cycle.  It required only about
15~seconds to switch between the two stars.
Over the four nights of
observations, we obtained 1036 individual spectra of \tar\ and 868
spectra of \nc, with a typical signal-to-noise ratio of $\sim 100$ near
2.2~\micron. 
Finally, we take calibration spectra approximately once per hour.
To account for flexure and
other changes over the night, important for the large SpeX instrument,
we record flats using an IR continuum lamp, and we record arc lamp
spectra for wavelength calibration.

\section{ANALYSIS}

In this section we discuss the details of the analysis process used to
interpret the observations.  This process consists of rejecting
outlying points (or `hot pixels'), extracting the spectra from the raw
data, and subtracting a suitable comparison spectrum to remove the
telluric features.   We search for changes in the resulting
`difference spectrum' that are synchronous with the secondary eclipse
and are therefore due to the spectrum of the planet.

\subsection{Spectral Extraction}

The first step in the analysis was the rejection of energetic particle
events and other intermittently bad pixels from the raw data.  This
was accomplished using a median filter over a set of eight raw data
frames, corresponding to two `abba' sets.  

For extracting the spectra from the raw data frames, we used the
Spextool software package (version 3.0), written by Mike Cushing and Bill Vacca for
analysis of SpeX data \citep{vacca03}.
The IDL widgets allow the user to enter an `ab'
pair of raw images, with the corresponding calibration files (flats
and arcs).  Given an individual `ab' pair, the program extracts the
sky background using the region of the slit between the `a' and `b'
object positions, and removes this from the extracted `a' and `b'
spectra.  Further, it checks for the residual sky background in the
pair-subtracted image, and removes it if necessary.  Currently, the
program outputs a standard extraction of the `a' and `b' spectra; that
is, it performs a sum over the spatial dimension to calculate the flux
at each wavelength point.  Spextool also performs a wavelength
calibration of the extracted spectra using the arc frames (recorded by
taking an image of an Argon lamp) as well as tabulated information on
telluric lines and comparing with the extracted spectra.  Finally,
Spextool corrects for slight detector non-linearities in the SpeX array.

\subsection{Corrections to the Extracted Spectra}

The spectra of each object for a given night are first interpolated
onto a constant and uniform
wavelength scale.  At this point we perform a second quality control
check to find and correct any remaining outlying points due to
uncorrected `hot pixels.'  For each
wavelength point we perform a median filter over the time series of all values
of that point for a given object during the night.  A point is
considered to be an outlier if the difference between the value of the
point and the median value is greater than $4 \sigma$, and it is then
replaced by the median value.  This process effectively removes
outlying points from the spectra, but it is only necessary to correct
about 0.3--0.4\% of
the total number of points for each object.

Next we remove from the stack any spectra that are clearly
discrepant.  The rejected spectra correspond in most cases to
observations that were listed in our notes as questionable, either
because of short-term
changes in terrestrial atmospheric conditions or poor focus of the
telescope.  The stacks of spectra for both objects now contain only
those spectra that will later be used in the calculation of the
difference spectra.

\begin{figure}
\plotone{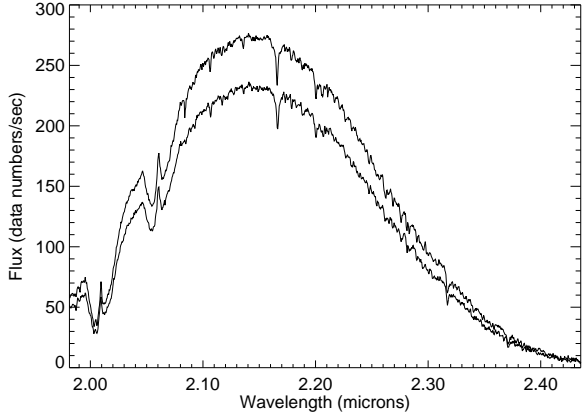}
\caption{Sample extracted spectra of \tar\ (lower, thick curve) and
\nc\ (upper, thin curve) from UT 27 September (near air mass of 1.27).
The comparison star has not yet been normalized for subtraction from
the \tar\ spectrum. The overall shape of the spectra is due to the
instrument response function.  Spectral features are primarily telluric.
\label{fig:spec_irtf}}
\end{figure}

With the individual spectra (again, of a given object) now
interpolated onto the same wavelength scale and cleaned of outlying
points, we then correct the data at each wavelength
point for air mass.  We fit a line to the log of
the intensity as a function of air mass, as described in
\citet[Equation 3]{r1}, although in this case, we correct to air mass
of unity.
In order to ensure that we do not remove the
effect of the planet from the \tar\ spectra, we experimented with
using the air mass
correction calculated for the comparison star to correct the \tar\
spectra.  We found that this did not have a large effect on the
resulting difference spectrum, showing that correcting the \tar\ and
\nc\ spectra separately did not introduce a bias in the resulting
residual spectrum.  An example of extracted spectra (only order 4,
$\sim$2.0--2.4~\micron) of \tar\ and \nc\
after the air mass correction is shown in Figure~\ref{fig:spec_irtf}.

We then calculate an average air 
mass-corrected spectrum of each object for that
particular night.  By comparing the average spectra of \tar\ and \nc\
we can identify stellar lines; at this resolution ($\lambda /
\Delta\lambda = 1500$), only
strong stellar lines are important.
Because of the large difference in the radial velocities of the two
stars, $|\Delta v| \simeq 55.2$~km~s$^{-1}$ \citep{nm02,wilson53},
a given stellar line will appear at a slightly different wavelength in
one star relative to the other.   These
show up clearly by subtracting the two average spectra.  We
remove them by linearly interpolating between the points on
either side of the line.  
In comparing
\tar\ and \nc, we found a total of 16 stellar lines, identified as Mg
and Si, as well as a single H line and a single Na line.

\subsection{Difference Spectra}
With the groups of spectra cleaned and adjusted for air mass, we are now ready to
consider individual spectra.  
First, we remove the known stellar lines from the individual spectra
in the same way that they were removed from the average spectra.
Note that at this point in the analysis, the two sets of
spectra (corresponding to the two stars) have been extracted and
interpolated onto separate wavelength scales.  
Next, an optimum shift value in wavelength is calculated (for both
the average as well as the individual spectra separately) by
minimizing the
standard deviation of the difference in intensity between a target and
a normalized comparison spectrum for each order separately. 
Using the calculated wavelength shift, we can then interpolate each
comparison spectrum onto the wavelength scale of each \tar\ spectrum.
This ensures that the
telluric absorption features line up in the average and individual
spectra of both objects.

Next, we calculate difference spectra by comparing
individual spectra of \tar\ and \nc.  As described, we
record spectra of \tar, typically in groups of
8, or two `abba' sets, and then switch to the comparison star.  The
subsequent observations of the comparison star are recorded at nearly
the same air mass as those of \tar.
We then calculate the normalized `difference spectrum' $d_i$ from
\begin{equation}
\label{eq:diff}
d_i = \frac{(t_i - f_i c_i)}{g_i \bar{c}}
\end{equation}
where $t_i$ represents a single spectrum of \tar, $c_i$ is the
corresponding comparison star spectrum, $f_i$ and $g_i$ are
normalization factors, and $\bar{c}$ is the average comparison star
spectrum for a given night of observations.  We calculate a difference
spectrum for each individual spectrum of \tar\ by subtracting the
corresponding subsequent comparison spectrum; for example, the first `a'
spectrum of \tar\ in a given `abba' set would be compared to the first
`a' spectrum in the subsequent comparison set.  The normalization
factors are used to place the comparison spectra (average and
individual) on the same scale as the spectra of \tar.  Although the
two stars are nearly the same brightness, we nevertheless expect that
they will not have exactly the same intensity levels, as seen in the
example in Figure~\ref{fig:spec_irtf}, due to slit
losses, variable atmospheric absorption, and changes throughout the
night.  Both normalization factors are calculated for the entire
spectrum, rather than for each order independently, to ensure that any
overall slope in the planetary spectrum is not removed.  The factors
are computed by enforcing the condition that the total of all orders
in the comparison spectrum is equal to the total of all orders in the
spectrum of \tar, as in
\begin{equation}
\label{eq:norm}
f_i = \frac{ \sum_{\lambda} t_i}{ \sum_{\lambda} c_i}
\end{equation}
and similarly for $g_i$.

This process effectively and completely removes the terrestrial
absorption lines that
dominate the object spectra.  Furthermore, the process also has two
desirable side-effects.
First, it removes
any time-varying changes in the detector response that may not have
been corrected by the flat-fielding.  Second, it removes many of the
variations in the continuum and line absorption that were not corrected by the
fit to air mass, because these variations may also be time-variable.
Note also that the resulting normalized
difference spectrum contains the candidate planetary spectrum in flux
units relative to the stellar spectrum, or the `contrast'
\citep{sudarsky03}.

\subsection{Averaging over Wavelength}
As described above, the individual difference spectra $d_i$ have been
calculated by subtracting an individual (i.e., at a given slit
position `a' or `b') \nc\ spectrum from the corresponding
individual \tar\ spectrum, as indicated by Equation~\ref{eq:diff}.
We then proceed to average the individual difference spectra $d_i$
over wavelength in order to improve the
signal-to-noise ratio.  We separate the spectra into
sections, or `bins', and average the points within each bin.  The bins
are defined specifically for each region, such that a bin boundary
does not fall on a telluric or stellar feature.  This ensures that we
do not introduce a bias in the bin average.  The width of each bin is
therefore variable but is typically $\sim 0.025$ \micron.

\begin{figure}
\plotone{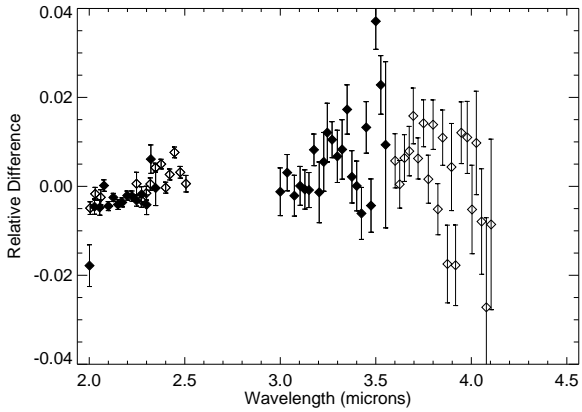}
\caption{Sample binned residual spectrum from UT 27 September,
corresponding to spectra shown in Figure~\ref{fig:spec_irtf}.  Five
different orders from the spectrograph
(three in the short wavelength region (1.9--2.5~\micron) and
two in the long wavelength region (3.0--4.1~\micron))
are shown as alternating filled and empty symbols.  Because of large
uncertainties resulting from the terrestrial background, the long
wavelength data are not considered further in this analysis.
\label{fig:binresid}}
\end{figure}

In looking at the series of binned difference spectra, we see that
most of them exhibit an overall slope.  The variation in intensity is
a slowly-varying function of wavelength, amounting to typically
1--5\% from 2--4~\micron; a slight slope is evident in the example
difference spectrum from UT 27 September shown in Figure~\ref{fig:binresid}.
We attribute this effect to image
motion on the slit due to guiding errors, coupled with the
$\lambda^{-0.2}$ wavelength dependence of
seeing \citep{linfield01}.  
Our interpretation is based on three considerations: 
1) the extreme slopes
are clearly related to image quality, as judged by the slit losses
through total intensity; 2) we constructed a simple model of the
process and found that we expect to see such slopes, based on the
wavelength dependence of seeing combined with guiding errors of
plausible magnitude; and 3)
the IRTF staff also see the effect (M.~Cushing, private communication,
2002).
We have investigated several methods for removing this
baseline effect, but we found that our results for the short
wavelength region (2--2.5~\micron) are robust and do not change
significantly based on whether a correction is made or on the details
of the correction.  This makes sense, given that the flux peak in the
planetary spectrum predicted by most models \citep[e.g.,][]{sudarsky03} is a
relatively sharp peak and therefore insensitive to the removal of a low-order
polynomial over the entire 2--4~\micron\ region.
We therefore present our results with no attempt to remove these gradual slopes.

Finally, the uncertainties have been carefully propagated through the
entire analysis.  At the beginning of the analysis, after the spectra
have been extracted from the raw frames using Spextool, we calculate
the noise level in each data set.  (Each data set consists of a
time-series of spectra of a given object for a given night of
observations.)  We take the error in each point to be the standard
deviation in the time series at each wavelength point independently.
This error value is propagated through the calculation of the
difference spectra and the wavelength binning using the standard error
propagation formulae.  Furthermore, we tested our analysis technique by
adding a synthetic planetary signal to the data and verifying that our 
algorithms could extract it with the correct amplitude.

\subsection{Fit to Eclipse Curve}
The final step in the analysis is the comparison of the processed
stack of residual difference spectra to the eclipse timing curve.  The
eclipse curve is calculated based on the predicted time of center of
secondary eclipse (corrected for the Earth-Sun light travel time), 
using known values of the period and time of center
eclipse from \citet{schultz} ($P = 3.5247542 \pm 0.0000044$~days
and $T_c = 2452223.896173 \pm  0.000086$~HJD)
and assuming a circular orbit (eccentricity $e=0$).  We also note that
updated ephemeris data ($P=3.5247501 \pm 0.0000004$~days and $T_c =
2452618.66888 \pm 0.00010$~HJD)
is available from \citet{wwo02}.
The eclipse curve is constructed in a simple way; zero represents
during eclipse, unity represents out of eclipse, and we perform a
linear interpolation between the ingress and egress points.  Then we
perform a linear least squares fit of the difference spectra to the
eclipse curve, at each wavelength bin.  The amplitude from the least
squares fit effectively produces a
final spectrum that represents the out-of-eclipse spectra minus the
in-eclipse spectra, and therefore represents the candidate planetary
spectrum.

Finally, we consider the possibility of a shift in the predicted time
of center of secondary eclipse, caused by a non-zero value of the
orbital eccentricity \citep{c03}.
The predicted times are based on the
assumption of zero eccentricity, which implies that the orbit is
circular and that the secondary eclipses occur exactly halfway between primary
eclipses.  However, the current value of the
eccentricity, based on a single-planet fit
to the Doppler velocity measurements, is
$e = 0.0281 \pm 0.0120$ with $\omega = 69.78 \pm 1.2^{\circ}$
(G.~Laughlin, private communication, 2003).
If this eccentricity is
taken at face value, it implies a shift in the timing of the secondary
eclipse of $\delta t = 31.4$ minutes, a slightly later eclipse.  
As shown below in Section~\ref{sec:res}, we find certain models for
the planetary atmosphere to be inconsistent with the data at $e=0$.
We have investigated the effect of changing the time of center eclipse by
as much as 120 minutes in either direction.
At no eclipse times do we find 
evidence for the rejected model spectra at full strength in the data.

\begin{figure}
\plotone{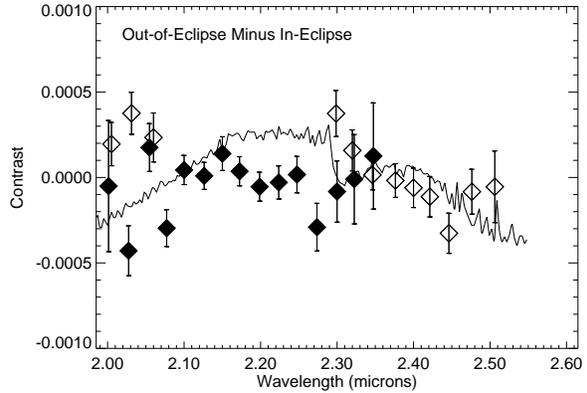}
\caption{Final difference spectrum, calculated by averaging the data
from the two `in-eclipse' nights.  This was obtained using 550
individual spectra of \tar, with an equal number of \nc\ spectra.  
The \citet{sudarsky03} baseline model is over-plotted at a
higher spectral resolution ($\lambda / \Delta \lambda \sim 2200$) than
the binned data.  
The offset between the model
and the data is not important, since we are comparing the shapes of
the two spectra; thus, the mean has been subtracted from the
data and model, respectively, for plotting purposes.
\label{fig:sw}}
\end{figure}

The final residual spectrum shown in Figure~\ref{fig:sw}
represents the resulting fit to the eclipse curve, averaged for both
of the in-eclipse nights, assuming zero orbital eccentricity.  We show 
only the result for the short wavelength region (1.9--2.5~\micron).
The larger error bars in the 3.0--4.0~\micron\ region (suggested by
Figure~\ref{fig:binresid}) are caused by
the terrestrial background and prevent a conclusive interpretation of
the planetary spectrum in this region.  Analysis of these
long-wavelength data is continuing but is beyond the scope of this paper.
Also plotted in Figure~\ref{fig:sw} is a
model spectrum by \citet{sudarsky03}---their baseline model for \tp;
this model is excluded by our data, as described in detail in Section~\ref{sec:res}.

We note that a flat line would be consistent with our data shown in
Figure~\ref{fig:sw}, implying that any model that does not exhibit the
peak shown in the \citet{sudarsky03} baseline model would escape detection in
the present analysis.  A flat line drawn through the data
overlaps 16 of the 27 points within the individual 1$\sigma$ error
bars, with a reduced chi-squared value of 2.35 (suggesting that the
error bars are slightly underestimated and smaller than the scatter in
the data).
Note that for a normal distribution of errors, 19 of 27 points would
be expected to fall within $\pm 1 \sigma$; therefore, our data are
roughly consistent with random deviations from a flat line.  We
verified that the errors in each wavelength bin are normally distributed.
In order to explore the range of models that are rejected by our data, we do some
diagnostic model calculations.

\section{MODEL CALCULATIONS}
Our model calculations are intended to help interpret the null result
for the planetary spectrum in the region between 2.0--2.5~\micron.  We
have implemented a simple spectral synthesis code to calculate model
spectra for \tp.  The algorithm consists of adopting a
temperature-pressure profile and computing the emergent thermal
emission spectrum.
We then scale the temperature-pressure profile in an ad hoc fashion
and re-compute the spectrum.
Although we do not enforce radiative equilibrium when scaling the
profile, the method is nonetheless consistent with our
intent, which is to provide a diagnostic of the planetary atmosphere.

The three fiducial temperature-pressure profiles, described below,
were used as input to the spectral
synthesis code and were calculated in radiative equilibrium 
using improvements on the method given by \citet{ss98} and \citet{sws00}.
We devised a simple way of scaling the fiducial temperature-pressure
profiles to see how these scalings changed the resulting spectrum.  The
prescription we used was
\begin{equation}
T_i^{\prime} = T_i + \gamma (T_i - T_0) +
T_{\mbox{\scriptsize{offset}}}, 
\label{eq:tscale}
\end{equation}
where $T_i$ is the temperature of layer $i$, and $T_0$ is the boundary
temperature, i.e., the temperature at small optical depth.
Two parameters control the shape of the synthetic profile:  $\gamma$ 
determines the temperature gradient, or atmospheric heating, and 
$T_{\mbox{\scriptsize{offset}}}$ is used to adjust the overall
temperature of the profile.
We characterize the shape of the resulting profile using
the temperature at optical depth unity for $\lambda = 2.15$~\micron\
(the center of the bandpass) and subtracting the boundary
temperature:
\begin{equation}
\Delta T = T(\tau_{\nu_c}\!=\!1) - T_0.
\label{eq:delt}
\end{equation}

Armed with a temperature-pressure profile, we continue by calculating
the continuous opacity due to collision-induced absorption (CIA) for
H$_2$-H$_2$ \citep{b02} and H$_2$-He \citep{j00}.  We include line opacities
for H$_2$O, CO, and CH$_4$.  The water lines are taken from the
extensive line database calculated by \citet{partsch} and compiled by
Kurucz,\footnote{\texttt{http://kurucz.harvard.edu/molecules/h2o/}}
the CO lines are from \citet{goorvitch94}, and the
CH$_4$ lines were obtained from HITRAN \citep{hitran}.  The relative
mixing ratios of the three species were determined using the simple
thermochemical equilibrium formulae provided by \citet{bs99}.  
We then compute emergent intensity as 
the formal solution to the radiative transfer
equation, with the source function equal to the Planck
function, from
\begin{equation}
\mu \frac{dI_\nu}{d\tau} = I_\nu - B_\nu(T).
\end{equation}
We then solve for $I_\nu$ and calculate the flux
density by integrating over $\mu$, as in
\begin{equation}
F_\nu = \int_\mu  \int_\tau B_\nu(T(\tau)) \,
e^{-\tau/\mu} \, d\tau \, d\mu. 
\label{eq:rtfinal}
\end{equation}
Finally, we calculate the contrast ratio
by dividing by a Kurucz model \citep{kurucz92} for \tar.  The
parameters used in the Kurucz model were $T=6000$~K, $\log g = 4.25$
(cgs), and [Fe/H]$=0.0$, as calculated by \citet{cody}
using the parameters in \citet{mazeh00}.

The assumption that the source function is given by the Planck
function amounts to ignoring scattering processes in the model atmosphere.
Only clouds and aerosols will produce significant scattering at this
wavelength.  Any attempt to specify the distribution of clouds in our
suite of diagnostic models would be problematic, since the structure
of these models has been perturbed in an ad hoc fashion.  Fortunately,
the amplitude of the flux peak at this wavelength is determined
primarily by the wavelength distribution of CO and H$_2$O opacity and the
temperature profile of the atmosphere.  

It is useful at this point to clarify our usage of the
\citet{sudarsky03} baseline model.
We use their tabulated flux
values\footnote{\texttt{http://zenith.as.arizona.edu/$\sim$burrows/}} for
the planet, divided by
the Kurucz model described above.  We also multiply by the area ratio
$0.016$, obtained by taking $R_p = 1.42 \, R_J$ \citep{cody} and $R_* =
1.146\, R_{\bigodot}$.
This yields approximately twice their tabulated contrast values, because
their contrast is phase-averaged (Sudarsky, private communication, 2002).
Using their temperature-pressure
profile as input, we calculate a flux peak of similar but slightly
larger magnitude with our simplified spectral synthesis code, and the
differences could be explained by the fact that we ignore their cloud
opacities in our calculation.  Note that our result is based on the
\citet{sudarsky03} baseline flux divided by the Kurucz model, \emph{not} on our
re-calculation of the spectrum.

\begin{figure}
\plotone{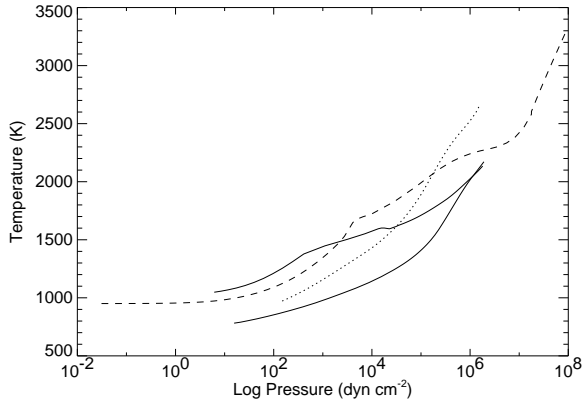}
\caption{Temperature-pressure profiles of the four fiducial cases.
The dashed curve is from the \citet{sudarsky03} baseline model, the thick solid
curve is the cloudy $f=1$ case, the thin lower curve is the
cloudless $f=1$ case, and the dotted line is the cloudless $f=2$ case.
\label{fig:fidprof}}
\end{figure}

\begin{figure}
\plotone{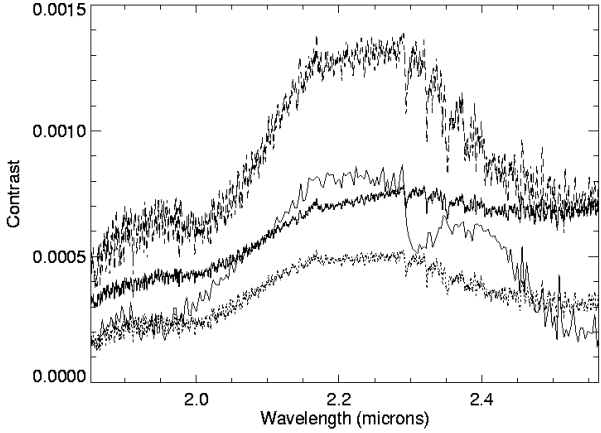}
\caption{Comparison of contrast for the four fiducial cases.
Uppermost dashed curve is the cloudless $f=2$ case.  The thick
middle curve is the cloudy $f=1$ case, and the lower dotted curve is the
cloudless $f=1$ case; these three
are the result of the simple spectral synthesis
calculations, plotted at a spectral resolution of 1500.  The thin,
solid curve is the \citet{sudarsky03} baseline result for the planetary flux
divided by the Kurucz model (see text).
\label{fig:fluxcomp}}
\end{figure}

The assumed redistribution of incident stellar radiation is also
different among the models, and we consider this factor in comparing
with our data.
Recalling the equation for the effective temperature of the planet
\citep{guillot96}, we have
\begin{equation}
\label{eq:teff}
T_{\mbox{\scriptsize{eff}}} = T_* \sqrt{\frac{R_*}{2D}} \left[ f(1-A) \right]^{1/4},
\end{equation}
where $T_*$ is the effective temperature of the star, $R_*$ is the
stellar radius, $D$ is the orbital radius of the planet, and $A$ is
the Bond albedo.  The parameter $f$ quantifies the redistribution of
incident stellar radiation; the $f=1$ case represents the situation in
which the incident radiation is evenly redistributed over the entire
planet, while the $f=2$ case represents the one in which the incident
radiation is completely absorbed and re-emitted on the day side of the
planet.  The \citet{sudarsky03} baseline model assumes $f=2$ (or equivalently,
$f=0.5$ using their definition of $f$ \citep{saumon02}).  As noted, our first case for
comparing with the observational results is the baseline \citet{sudarsky03}
flux divided by the Kurucz model.  We choose three other cases as
fiducial models for comparison with the data.  These have been
selected in order to test a wide range of the possible physical
conditions in the atmosphere.  We choose two $f=1$ models based on the
formulation by \citet{ss98, sws00}, one with
no clouds and one with cloud opacity due to MgSiO$_3$, Fe, and
Al$_2$O$_3$.  Since the \citet{sudarsky03} baseline model considers cloud
opacities and assumes $f=2$, we further select a cloudless model with
$f=2$ (based on the \citet{ss98, sws00} formulation).
The temperature profiles of
these four cases are shown in Figure~\ref{fig:fidprof}, and
the results for the four flux peaks (the \citet{sudarsky03} baseline result as
well as the cloudless and cloudy fiducial cases) 
are shown in Figure~\ref{fig:fluxcomp}.

\section{RESULTS AND INTERPRETATION}
\label{sec:res}
We now fit these four modeled flux peaks
to the data using linear least squares, as in
\begin{equation}
r_i = a + b m_i,
\label{eq:lsfit}
\end{equation}
where $r_i$ represents the resulting difference spectrum (e.g.,
Figure~\ref{fig:sw}), $m_i$ represents the contrast ratio from a
given model, $i$ represents the wavelength bin, and $a$ and $b$ are
the coefficients of the fit.
The slope of this fit, $b$, which we call the `model amplitude,' therefore
represents the `amount' of the model signal that appears in the data.  
A model amplitude of unity would mean that the model clearly fits the
data (depending on the errors),
while a value of zero represents no correlation between the
model and the data.  
Note that the $y$-offset
between the model and the data is irrelevant, since we are searching
only for the evidence of the shape of a given model within the data.
The results for the four fiducial models are shown in Table~\ref{tbl:mares}.
For each model, we indicate the model amplitude $b$, the uncertainty
in $b$, and the reduced chi-squared of the fit.  We further list the
reduced chi-squared for the case in which we \emph{force} the model amplitude
to be unity; that is, we assume that the model appears in the data at
full strength.
For all four fiducial models, the slope of the fit is roughly zero,
suggesting that none of them fit the data.
However, the rejection is much stronger for the both $f=2$ models, as
seen by the reduced chi-squared values for forcing the model amplitude
to be unity.
These high values indicate that it is extremely unlikely that the
$f=2$ models fit the data.
Given that the observations and analysis are highly differential, 
we are furthermore confident that a peak in the planetary spectrum is not being
fortuitously canceled by some systematic error in the data.

\begin{figure}
\plotone{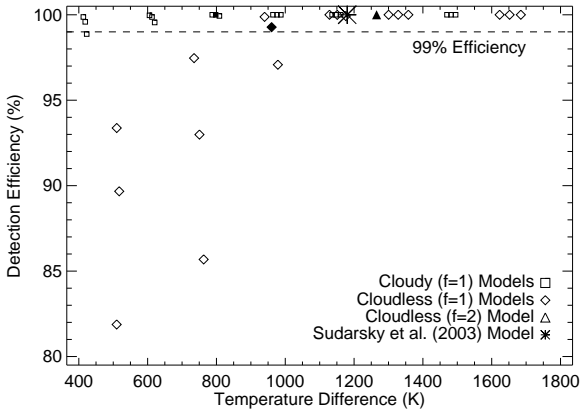}
\caption{Detection efficiency $Q$ for a suite of
scaled models vs.\ $\Delta T$, which represents the difference between
the temperature at optical depth unity at the center of the bandpass
and the boundary temperature (see Equation~\ref{eq:delt}).
See Tables~\ref{tbl:ncmodels} and~\ref{tbl:cmodels}
for details of each scaled profile.  The filled symbols, along with
the asterisk, represent the four
fiducial cases described in Table~\ref{tbl:fidmod}.
\label{fig:rej} }
\end{figure}

The least squares fit described above is difficult to interpret for
models with small amplitude flux peaks, as in some of our scaled
models (see Equations~\ref{eq:tscale} and~\ref{eq:delt}).
It is also desirable to explore another way to quantify the degree to
which the four specific models are rejected by the data.
We use a Monte Carlo technique to accomplish this.
To quantify the statistical significance
of the rejection of any particular
model, we constructed `fake data' by adding
normally-distributed noise to the modeled contrast values, with
$\sigma$ equal to the size of the error bar in each bin of
Figure~\ref{fig:sw}.
We created 10$^5$ fake data sets for each model and then fit each one
to the original model using linear least squares (in the same way in
which the real data were fit to the model to calculate the model
amplitude).  Next, we determine the number of resulting
model amplitudes which are greater than the value obtained by fitting
the original model to the actual data; this is a liberal criterion for
concluding that the model is consistent with a given fake data set.  
The percentage of fake data sets consistent with a given model,
using this criterion, is the `detection
efficiency', $Q$.  A high value of $Q$, say $>99\%$, implies that a
given model consistently produced at least a minimal signature in the fake data.
The results for the four specific models discussed thus far are shown
in Table~\ref{tbl:fidmod}, and are also plotted in
Figure~\ref{fig:rej} (filled symbols).
A model with a high detection efficiency using the Monte Carlo
technique implies that it is detectable.  Thus, the two $f=2$
models---having model
amplitudes near zero and high detection efficiencies---are again
concluded to be rejected.
These results are consistent with the chi-squared analysis described above.

We tested our suite of scaled models, with the prescription
given by Equation~\ref{eq:tscale}, using this technique.  The
parameters and results of these models, as well as the model amplitudes, are shown in
Table~\ref{tbl:ncmodels} for the cloudless $f=1$ cases and in
Table~\ref{tbl:cmodels} for the cloudy $f=1$ cases.  The results of
$Q$ vs.\ $\Delta T$ are plotted in
Figure~\ref{fig:rej} for both sets of models.  
We reject a model if it has a value of $Q>99\%$
and also has a model amplitude near zero.  Looking at
Tables~\ref{tbl:ncmodels} and~\ref{tbl:cmodels}, the rejected models
correspond to the steeper T/P profiles, with $\gamma \geq 1.25$.
Thus, we can see that the
temperature profiles with a higher temperature gradient,
causing a larger peak near
2.2~\micron\ in the emergent spectrum, can be eliminated as
inconsistent with our observations.  The more isothermal profiles, on
the other hand, are consistent with the data.  
Figure~\ref{fig:rej}
suggests that only models with $\Delta T \lesssim 1100$~K are
consistent with our data.  
Although the rejection limit also depends on
$T_{\mbox{\scriptsize{offset}}}$ from Equation~\ref{eq:tscale},
it is strongly dependent on $\Delta T$.

\section{CONCLUSION}

Our results are extremely sensitive to the temperature-pressure
profile, which is as expected, 
given that the Planck function varies strongly with temperature in
this wavelength region.
Besides the shape of the profile, there are a variety of other
possible processes in the planetary atmosphere that could render the
planet undetectable with these observations.
For example, the broad peak in the spectrum near 2.2~\micron\ is caused by the
large number of weak water lines on either side of this feature.
Absorption by these lines lowers the continuum level on either side,
thus creating the peak.  One obvious process by which the peak would not
appear is a lowering of the water abundance in the planetary
atmosphere.  We tested this idea and
found that only when the water abundance was 
lowered by a factor of 10 did the peak become small enough to
escape detection using our method.  This value for the water abundance
($\sim 9.8 \times 10^{-5}$) seems unrealistically small.

We suggest three ways in which the 2.2~\micron\ flux peak would be
lowered sufficiently to be consistent with our observations.  The
first is the efficient redistribution of the incident radiation over the entire
planet, which can be achieved through atmospheric dynamics, i.e.,
winds.  Depending on how efficiently the incident radiation is
redistributed, the effect could be a temperature asymmetry between the
`day' and `night' sides of the planet.
\citet{sg02} suggest that the day-night temperature
difference may be as much as 300~K; even a difference of this
magnitude would lead to strong winds of 1~km~s$^{-1}$ or more.  The
$f=2$ situation would tend to steepen the T/P profile in the
atmosphere, since the strong stellar irradiation would have to be
absorbed relatively deep in the atmosphere and re-radiated.
The two specific $f=2$ models considered above, which assume that the heat is
absorbed and re-radiated only on the day side, are inconsistent with our data (as
shown in Table~\ref{tbl:mares}).  
This suggests that the large temperature asymmetry scenario may
not be viable and that some sort of transport mechanism may be at
work to move the incident energy from the hot to the cold side.
We caution that we have observed only
two secondary eclipses of \tp; recent simulations by
\citet{cho} indicate that the temperature asymmetry between the two
hemispheres may be time-variable.  This effect can be so strong as to
render the day side to be colder than the night side at times.

A second mechanism by which the 2.2~\micron\ flux peak could be lowered
is additional opacity in the upper atmosphere, which would flatten the T/P
profile and thus reduce the relative strength of the peak.  For
example, the T/P profiles shown by \citet{barman01} are
nearly isothermal, and will certainly produce 2~\micron\ spectra
consistent with our observed limit.  Barman (2003, private
communication) attributes this behavior to their treatment of the
opacity, specifically the inclusion of strong gaseous opacity due to
metals, which differs from the opacity treatment in other models.
However, consistency with our data does not necessarily
indicate that a given model is correct, and we leave it to the
theorists to sort out the differences among competing models.
We suggest that a high-altitude cloud (first suggested by \citet{ss00}
and favored by
\citet{cb02} to explain their low sodium result) would
also make the atmosphere more isothermal if the cloud is sufficiently
absorbing and vertically extended.  Note, however, that the \citet{cb02} sodium result
requires placing such a cloud at the limb of the planet, whereas our
results require the clouds to be broadly distributed on the day
side.

A high cloud might also affect the emergent spectrum through
a third mechanism.  For a sufficiently high cloud, the atmosphere above the cloud
deck would be thin, giving a small column density in which a spectral
signature could be formed.  Very little incident radiation would
penetrate an optically thick cloud, meaning that the temperature structure would be
dominated by the blackbody reflection or emission of the cloud itself.  The emergent
spectrum would be determined by the small part of the atmosphere above
the cloud deck, rather than the details of the T/P profile.

Finally, we note that our results have been obtained using the
3-meter IRTF, indicating that valuable observations of extrasolar
planets, leading to new information on the atmospheres of these
objects, can be conducted using modest-aperture telescopes from the ground,
given favorable observing conditions and careful analysis.

\acknowledgments
This paper is based upon observations at the IRTF, supported by the National
Aeronautics and Space Administration under Cooperative Agreement no.\
NCC 5-538 issued through the Office of Space Science, Planetary
Astronomy Program.
We thank Dave Sudarsky for providing us with his temperature-pressure
profile and for clarifying discussions of these results.
We thank
the IRTF staff, in particular Mike Cushing for his lengthy discussions
of the Spextool extraction software and John Rayner for his insights into
the SpeX instrument.
Finally, we thank the referee for helpful comments and suggestions.
Jeremy Richardson is supported in part by a NASA Graduate Student Researchers
Program Fellowship, funded by the Office of Space Science at NASA
Headquarters (grant number NGT5-50273).  Other aspects of this research
were supported by the NASA Origins of Solar Systems Program.  Sara
Seager is supported by the Carnegie Institution of Washington.

\clearpage

\bibliographystyle{apj}
\bibliography{/home/jeremy/thesis/main}

\clearpage

\begin{deluxetable}{lrrr|r}
\tabletypesize{\scriptsize}
\tablecaption{Results of the least squares fit of the data to each of
  the four fiducial models, indicating the slope (model amplitude) $b$
  and the uncertainty in $b$, as well as the reduced chi-squared for
  the fit.  The rightmost column indicates the reduced chi-squared
  value for the case in which we force the slope to be unity.
\label{tbl:mares}}
\tablewidth{0pt}
\tablehead{
\colhead{Model} & \colhead{$b$} & \colhead{$\sigma_b$} &
\colhead{$\bar{\chi}^2$} & \colhead{$\bar{\chi}^2 (b=1)$} }

\startdata
Cloudless (f=1)	& 0.02 & $\pm$ 0.29 & 2.35 & 2.85 \\ 
Cloudless (f=2) & 0.03 & $\pm$ 0.10 & 2.34 & 6.43 \\
Cloudy (f=1) & -0.25 & $\pm$ 0.24 &  2.30 & 3.43 \\ 
Cloudy (f=2)\tablenotemark{*} & -0.01 & $\pm $ 0.14 & 2.35 & 4.66 \\

\enddata
\tablenotetext{*}{As calculated by dividing the flux from
  the \citet{sudarsky03} baseline model by the
Kurucz model for the stellar flux; 
see Figure~\ref{fig:fluxcomp} (thin, middle curve).}
\end{deluxetable}

\clearpage

\begin{deluxetable}{lrr}
\tabletypesize{\scriptsize}
\tablecaption{Results from the Monte Carlo technique for the four
  fiducial models.  The value of $\Delta$T from Equation~\ref{eq:delt}
  is shown, as well as the value $Q$ describing the efficiency of the
  detection using the Monte Carlo technique,
 as plotted in Figure~\ref{fig:rej}.
\label{tbl:fidmod}}
\tablewidth{0pt}
\tablehead{
\colhead{Model} &
\colhead{$\Delta$T (K)} & \colhead{Q (\%)} }

\startdata
Cloudless (f=1)	& 960.36 & 99.3\\ 
Cloudless (f=2) & 1265.38 & 100.0\\
Cloudy (f=1) & 797.92 & 100.0\\ 
Cloudy (f=2)\tablenotemark{*} & $\sim$ 1180 & 100.0\\

\enddata
\tablenotetext{*}{As calculated by dividing the flux from
  the \citet{sudarsky03} baseline model by the
Kurucz model for the stellar flux; 
see Figure~\ref{fig:fluxcomp} (thin, middle curve).}
\end{deluxetable}

\clearpage

\begin{deluxetable}{r|rr|rr|r}
\tabletypesize{\scriptsize}
\tablecaption{Results for scaled cloudless $f=1$ profiles.
  The first column is the value of $\Delta T$ from
  Equation~\ref{eq:delt}, and the scaling parameters are defined in
  Equation~\ref{eq:tscale}.  The model amplitude $b$ and uncertainty
  $\sigma_b$ are also shown.
The value $Q$ represents the detection efficiency using the Monte Carlo
technique, as plotted in Figure~\ref{fig:rej} (diamonds).
\label{tbl:ncmodels}}
\tablewidth{0pt}
\tablehead{
\colhead{$\Delta$T (K)} & \colhead{$\gamma$} & \colhead{T$_{\mbox{offset}}$}
& \colhead{b} & \colhead{$\sigma_b$} & \colhead{Q (\%) }}

\startdata
  517.42 &   0.50 &    0 & -1.49 &  1.35 &  89.7\\
  510.21 &   0.50 &  100 & -0.78 &  0.81 &  93.4\\
  510.00 &   0.50 & -100 & -1.84 &  2.05 &  81.9\\
  750.58 &   0.75 &    0 & -0.13 &  0.53 &  93.0\\
  734.92 &   0.75 &  100 & -0.13 &  0.40 &  97.5\\
  762.85 &   0.75 & -100 & -0.16 &  0.75 &  85.7\\
  960.36 &   1.00 &    0 &  0.02 &  0.29 &  99.3\\
  939.94 &   1.00 &  100 & -0.00 &  0.24 &  99.9\\
  978.55 &   1.00 & -100 &  0.05 &  0.36 &  97.1\\
 1151.05 &   1.25 &    0 &  0.04 &  0.18 & 100.0\\
 1128.48 &   1.25 &  100 &  0.02 &  0.16 & 100.0\\
 1175.98 &   1.25 & -100 &  0.06 &  0.21 &  99.9\\
 1328.16 &   1.50 &    0 &  0.04 &  0.13 & 100.0\\
 1300.31 &   1.50 &  100 &  0.03 &  0.11 & 100.0\\
 1357.76 &   1.50 & -100 &  0.05 &  0.14 & 100.0\\
 1652.07 &   2.00 &    0 &  0.03 &  0.07 & 100.0\\
 1623.06 &   2.00 &  100 &  0.03 &  0.07 & 100.0\\
 1684.67 &   2.00 & -100 &  0.04 &  0.08 & 100.0\\
\enddata
\end{deluxetable}

\clearpage

\begin{deluxetable}{r|rr|rr|r}
\tabletypesize{\scriptsize}
\tablecaption{Results for the scaled cloudy $f=1$ profiles.  The same
  quantites as described in Table~\ref{tbl:ncmodels} are shown.  The values for
  the detection efficiency $Q$ are also plotted in
  Figure~\ref{fig:rej} (squares).
\label{tbl:cmodels}}
\tablewidth{0pt}
\tablehead{
\colhead{$\Delta$T (K)} & \colhead{$\gamma$} & \colhead{T$_{\mbox{offset}}$}
& \colhead{b} & \colhead{$\sigma_b$} & \colhead{Q (\%)} }

\startdata
  418.63 &   0.50 &    0 & -0.83 &  0.49 &  99.6\\
  413.15 &   0.50 &  100 & -0.65 &  0.38 &  99.9\\
  422.91 &   0.50 & -100 & -1.14 &  0.65 &  98.9\\
  612.63 &   0.75 &    0 & -0.46 &  0.33 &  99.9\\
  605.37 &   0.75 &  100 & -0.38 &  0.27 & 100.0\\
  620.46 &   0.75 & -100 & -0.58 &  0.42 &  99.6\\
  797.92 &   1.00 &    0 & -0.25 &  0.24 & 100.0\\
  787.74 &   1.00 &  100 & -0.22 &  0.20 & 100.0\\
  808.55 &   1.00 & -100 & -0.30 &  0.28 &  99.9\\
  976.03 &   1.25 &    0 & -0.14 &  0.18 & 100.0\\
  964.24 &   1.25 &  100 & -0.13 &  0.16 & 100.0\\
  987.11 &   1.25 & -100 & -0.16 &  0.20 & 100.0\\
 1147.48 &   1.50 &    0 & -0.08 &  0.14 & 100.0\\
 1136.11 &   1.50 &  100 & -0.08 &  0.12 & 100.0\\
 1160.71 &   1.50 & -100 & -0.09 &  0.15 & 100.0\\
 1481.52 &   2.00 &    0 & -0.03 &  0.09 & 100.0\\
 1470.01 &   2.00 &  100 & -0.03 &  0.08 & 100.0\\
 1495.06 &   2.00 & -100 & -0.03 &  0.10 & 100.0\\
\enddata
\end{deluxetable}

\clearpage

\end{document}